\documentclass[prl,english,superscriptaddress,twocolumn]{revtex4-1}

\usepackage{amsmath,amsfonts,amssymb}
\usepackage{graphicx,float,calc}
\usepackage[breaklinks,colorlinks,urlcolor=blue,citecolor=blue,linkcolor=blue]{hyperref}

\usepackage{xcolor}

\begin{document}

\title{
Quantum Simulation of a Hyperbolically Driven Quantum System with Superconducting Circuits
}

\author{Han-xin Che}
\thanks{These authors contributed equally to this work.}
\affiliation{National Laboratory of Solid State Microstructures, School of Physics, Nanjing University, Nanjing 210093, China}
\affiliation{Shishan Laboratory, Suzhou Campus of Nanjing University, Suzhou 215000, China}

\author{Yangyang Yu}
\thanks{These authors contributed equally to this work.}
\affiliation{National Laboratory of Solid State Microstructures, School of Physics, Nanjing University, Nanjing 210093, China}
\affiliation{Shishan Laboratory, Suzhou Campus of Nanjing University, Suzhou 215000, China}

\author{Tao Zhang}
\affiliation{National Laboratory of Solid State Microstructures, School of Physics, Nanjing University, Nanjing 210093, China}
\affiliation{Shishan Laboratory, Suzhou Campus of Nanjing University, Suzhou 215000, China}

\author{Wen Zheng}
\affiliation{National Laboratory of Solid State Microstructures, School of Physics,
    Nanjing University, Nanjing 210093, China}
    \affiliation{Shishan Laboratory, Suzhou Campus of Nanjing University, Suzhou 215000, China}
\affiliation{Hefei National Laboratory, Hefei 230088, China}

\author{Hai-Feng Yu}
\email{hfyu@baqis.ac.cn}
\affiliation{Beijing Key Laboratory of Fault-Tolerant Quantum Computing, Beijing Academy of Quantum Information Sciences, Beijing 100193, China}
\affiliation{Hefei National Laboratory, Hefei 230088, China}

\author{Xinsheng Tan}
\email{tanxs@nju.edu.cn}
\affiliation{National Laboratory of Solid State Microstructures, School of Physics,
    Nanjing University, Nanjing 210093, China}
    \affiliation{Shishan Laboratory, Suzhou Campus of Nanjing University, Suzhou 215000, China}
\affiliation{Hefei National Laboratory, Hefei 230088, China}

\author{Yang Yu}
     \email{yuyang@nju.edu.cn}
\affiliation{National Laboratory of Solid State Microstructures, School of Physics,
    Nanjing University, Nanjing 210093, China}
    \affiliation{Shishan Laboratory, Suzhou Campus of Nanjing University, Suzhou 215000, China}
\affiliation{Hefei National Laboratory, Hefei 230088, China}

\begin{abstract} 
Hyperbolic geometry hosts a plethora of exotic quantum phenomena ranging from holographic duality to novel topological phases. 
However, investigating these effects experimentally remains challenging due to the Euclidean constraints of natural crystalline materials. 
While recent synthetic lattices have achieved static realizations, they often lack the dynamical tunability required to explore the full geometric landscape. 
Here, we overcome these limitations by implementing a quantum simulation of hyperbolic geometry within the parameter space of a superconducting quantum circuit. 
We construct a Hyperbolically Driven Quantum System (HDQS) by mapping the hyperbolic properties onto the dynamically controllable Hamiltonian parameters. 
By probing the system's response under geometric quantum driving, we elucidate how the intrinsic geometry of the parameter space dictates observable physical quantities. 
Furthermore, we extend the theoretical framework to realize ensemble simulation. 
This generalization enables us to harness the fast coherent control of superconducting qubits to capture essential geometric characteristics, establishing a versatile platform for exploring quantum dynamics on hyperbolic manifolds.
\end{abstract}

\maketitle

\emph{Introduction.---}
Geometry lies at the heart of modern physics, providing the fundamental framework for understanding the interactions of matter and fields\cite{Torma2022NRP, Cisowski2022RevModPhys, Torma2023PRL, liu2025quantumgeometry, Frankel_2011}. 
    While our intuition is grounded in Euclidean geometry, hyperbolic geometry—characterized by constant negative curvature—has emerged as a cornerstone in theoretical physics, bridging diverse frontiers from high-energy physics\cite{Swingle2012, Salvatori2018hyperbolic} to quantum information\cite{Breuckmann2016, Brody2001geometric}. 
AdS/CFT correspondence\cite{Aharony2000} suggests that a quantum theory of gravity\cite{Maldacena1999} in the bulk of a hyperbolic space is equivalent to a conformal field theory\cite{DiFrancesco1997} on its boundary, sparking intense interest in simulating hyperbolic geometries to probe the mysteries of black holes\cite{Maldacena1999, Luminet1995topology} and quantum entanglement\cite{Ryu2006} in a controlled laboratory setting.

Beyond high-energy physics, the introduction of hyperbolic geometry into condensed matter systems has unveiled a plethora of exotic phenomena absent in conventional Euclidean lattices. Systems defined on hyperbolic manifolds exhibit unique spectral properties, such as macroscopic degeneracy of eigenstates (flat bands)\cite{Maciejko2021} and a distinctive bulk-boundary correspondence\cite{Yu2020} arising from the exponential growth of volume relative to surface area\cite{Boettcher2020}. 
These features host promising platforms for realizing novel topological phases of matter\cite{Chiu2016} and robust quantum error correction codes\cite{Breuckmann2021}. 
However, investigating these phenomena experimentally poses a significant challenge: natural crystalline materials are strictly constrained by Euclidean crystallography, making the direct synthesis of “hyperbolic crystals” in real space remarkably difficult\cite{boettcher2020hyperbolic}. 
While recent breakthroughs have realized hyperbolic lattices using coplanar waveguide resonators\cite{Xu2025} and topolectrical circuits\cite{Lee2018, Lenggenhager2022}, these static implementations often lack the dynamical tunability required to explore the full landscape of hyperbolic topology\cite{Urwyler2022, Huang2024}.

To overcome the constraints of real-space geometry, quantum simulation\cite{Georgescu2014, Daley2022} in parameter space\cite{Gritsev2012, Hu2018} offers a powerful alternative. 
By mapping the geometric properties of a manifold onto the Hamiltonian parameters of a controllable quantum system, one can investigate the intrinsic geometry—characterized by the quantum metric\cite{Provost1980} and Berry curvature\cite{Berry1984}—without the need for physical spatial curvature. 
Superconducting quantum circuits\cite{CircuitQED}, with their  controllability, stand out as an ideal platform for this approach\cite{Schroer2014, Tan2019, Roushan2014}. 
The ability to dynamically drive system parameters allows for the engineering of effective Hamiltonians that traverse specific trajectories in parameter space, thereby simulating the transport of quantum states and exploring the topology of hyperbolic space\cite{Bienias2022}.

\begin{figure}
    \centering
    \includegraphics[width=8cm]{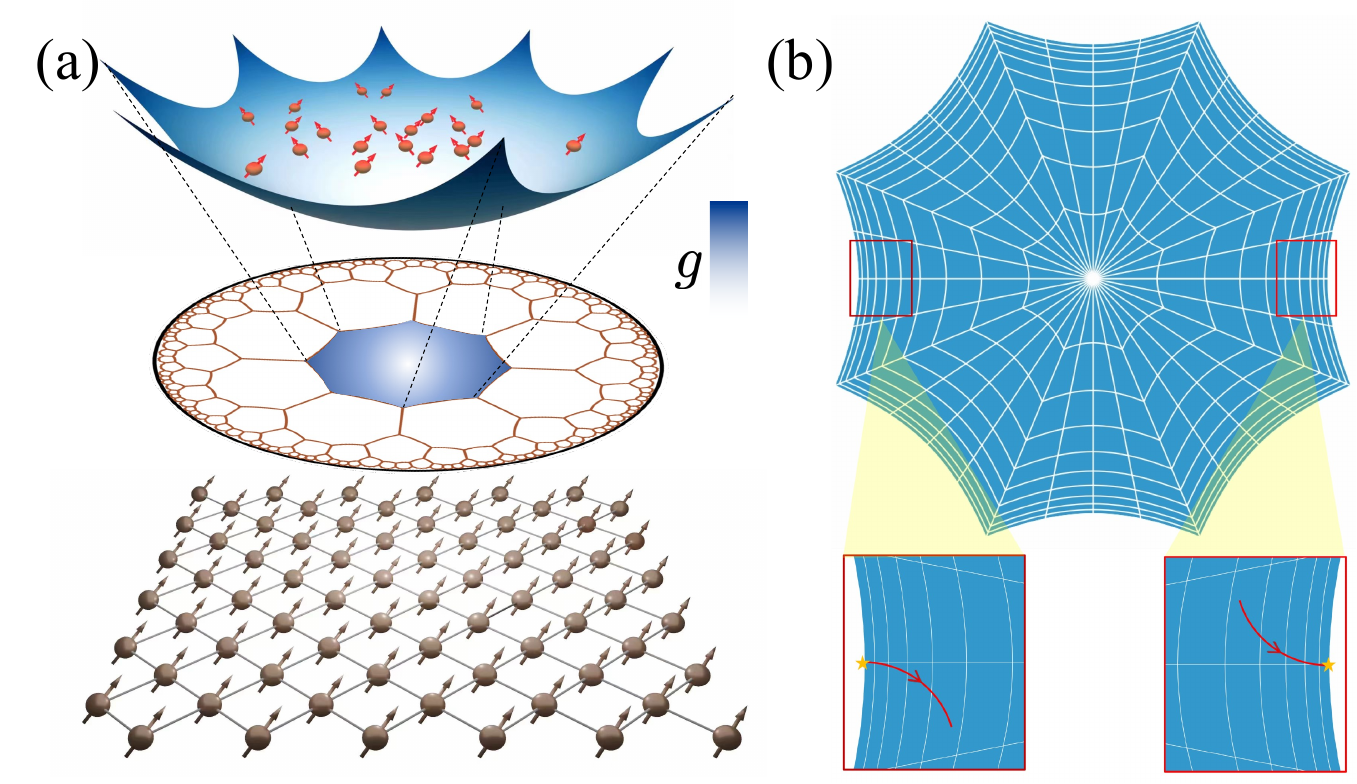}
    \caption{Simulation of quantum dynamics on a hyperbolic parameter manifold. 
    (a) The top panel illustrates an ensemble of particles evolving within the hyperbolic parameter space. 
    The central part displays the Bolza surface, a quotient space of the Poincaré disk, where spatial distortion increases with distance from the center.
    The bottom panel shows the superconducting quantum chip, where a one-to-one mapping is established between each physical qubit and a virtual particle.
    (b) On the Bolza surface, opposite edges are equivalent.
     When a particle crosses one boundary edge, it re-enters from the opposite side.}
    \label{fig:FiG0}
\end{figure}

In this work, we simulate a hyperbolically driven quantum system\cite{Wu2026GQD} using a superconducting quantum circuit platform. 
By investigating the system’s response under varying degrees of adiabaticity and ergodicity, we reveal how the geometry of the parameter space influences observable physical quantities.
Furthermore, we extend the theoretical framework to realize ensemble simulation. 
This extension not only leverages the parallel computing capabilities of superconducting quantum chips but also enables the efficient capture of geometric features, effectively transcending temporal constraints. 
Consequently, it offers novel insights and methodologies for investigating quantum geometry on exotic manifolds and advancing manifold-based quantum machine learning\cite{Stokes2020quantum, Manna2025arxiv, Krioukov2010PRE}.

\emph{Construction of a hyperbolically driven quantum system.---}
To simulate quantum dynamics on a hyperbolic manifold, we implement a parameter-space driving scheme where the system's Hamiltonian is modulated along trajectories on a curved geometry. As illustrated in Fig. \ref{fig:FiG0}(a), the top panel depicts the theoretical framework: an ensemble of classical particles evolving on a hyperbolic parameter space $\mathcal{M}$. The bottom panel shows the experimental realization on a superconducting quantum chip, establishing a one-to-one mapping between physical qubits and virtual particles.

Theoretically, a parametrically driven system is defined by the triple $(\mathcal{M}, \hat{H}(x), x(t))$. Here, we utilize the Bolza surface $\Sigma_2$—a closed Riemann surface with genus $g=2$ and constant negative curvature—as our parameter manifold $\mathcal{M}$. The Bolza surface is constructed as the quotient of the Poincaré disk $\mathbb{D}=\{z \in \mathbb{C} \mid |z|<1\}$ under the action of a Fuchsian group. 
The quotient map makes its boundary cyclic, Fig. \ref{fig:FiG0}(b).
The metric is given by $g_{ij}=4\delta_{ij}/(1-|z|^2)^2$.

In this article, we synthesize a two-level Hamiltonian $\hat{H}(z)=\boldsymbol{\Omega}(z)\cdot\boldsymbol{\Lambda}$ , parameterized by the coordinate $z$ on the manifold, where $\boldsymbol{\Lambda}$ are spin operators. 
The parameter $\epsilon$ controls the topological phase of the system.
Experimentally, we set $\Omega$=20MHz.

\begin{figure}
    \centering
    \includegraphics[width=9cm]{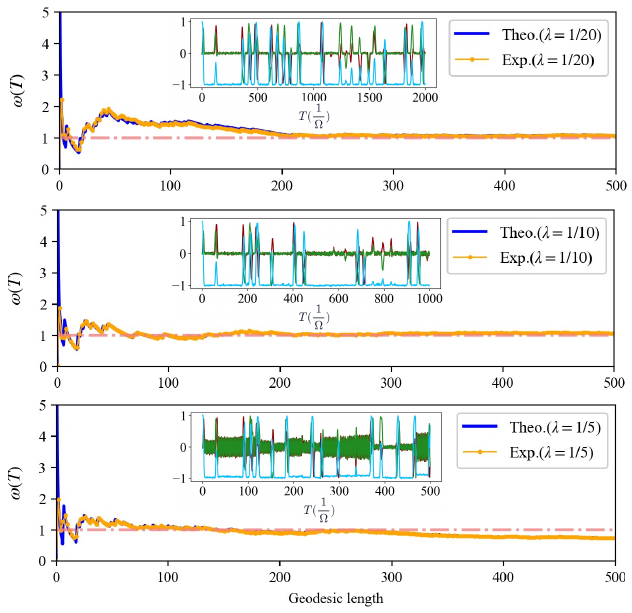}
    \caption{Quantum tomography and topological response.
      Quantum state tomography experimentally shows the trajectory of the quantum state on the Bloch sphere during the first one-fifth of the evolution time(500, 250, 125$\mu s$), corresponding to the classical particle with $\lambda = 1/20$, $\lambda = 1/10$, and $\lambda = 1/5$, respectively.
      The red, green and blue solid line represent the value of $\langle\sigma_x\rangle$, $\langle\sigma_y\rangle$ and $\langle\sigma_z\rangle$.
     At a suitable driving speed, the topological response converges to the first Chern number with increasing geodesic length.}
     \label{fig:FiG1}
\end{figure} 

\emph{Topological response and adiabatic limits.---}
A key question is whether the global topological properties of the quantum system can be recovered by driving the system along specific trajectories. While local driving is generally insufficient, the Hopf–Anosov theorem guarantees that geodesic flow on a closed, negatively curved surface is ergodic. This implies that a particle moving along a geodesic will eventually explore the entire phase space, allowing for the extraction of global invariants.

\begin{figure*}[t]
    \centering
    \includegraphics[width=15cm]{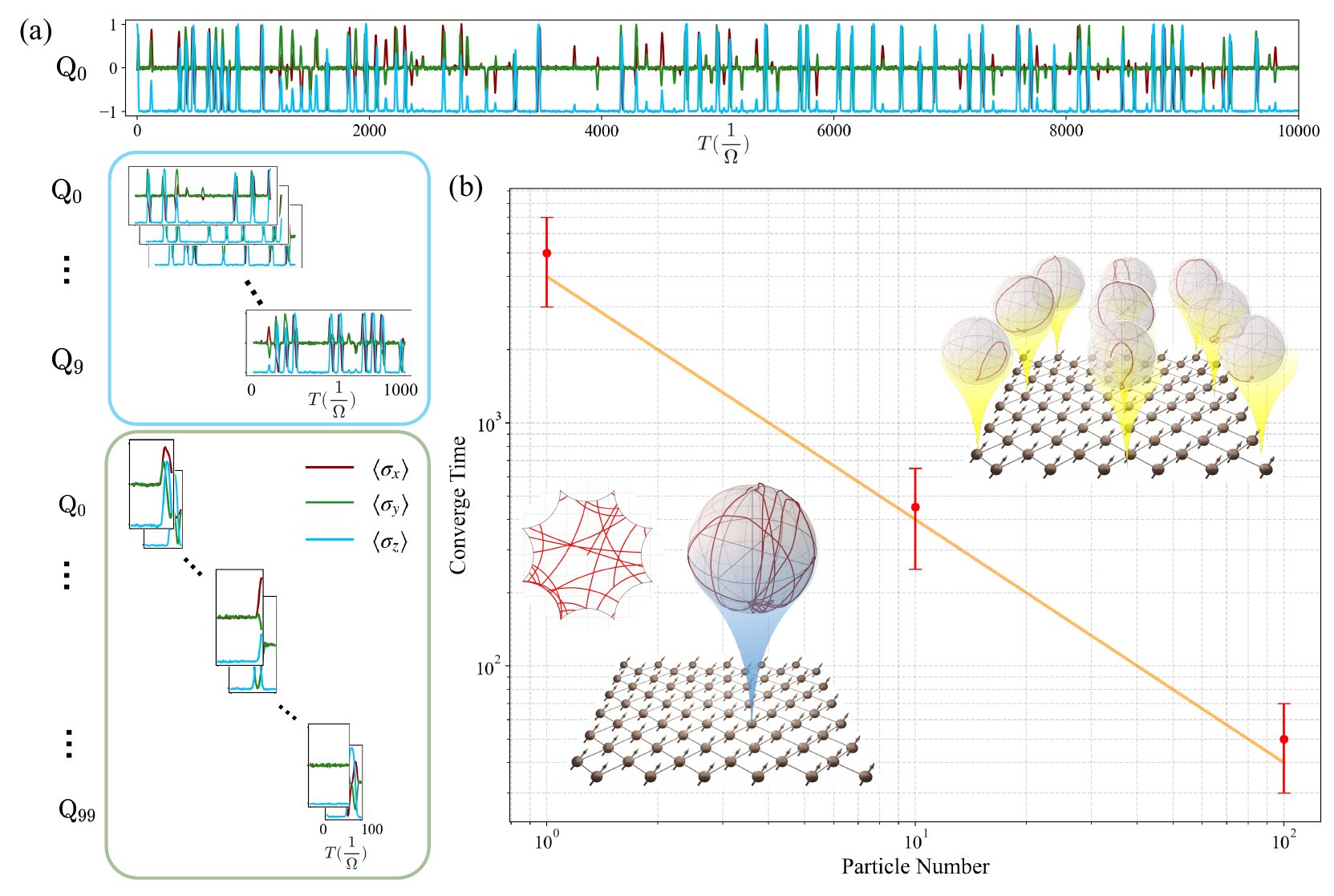}
    \caption{Parallel acceleration and quantum state trajectories.
    (a) From top to bottom: time evolution(500, 50, 5$\mu s$) of quantum state tomography results for particle numbers $N = 1, 10,$ and $100$, respectively.
    (b) Convergence time required to extract the topological number from the dynamical response.
    Bottom-left: Evolution of the quantum state on the Bloch sphere and the particle trajectory on the hyperbolic manifold for $N = 1$. For brevity, only the initial short time interval is displayed.
    Top-right: Evolution of a subset of quantum states on the Bloch sphere for $N = 100$.
    }
    \label{fig:FiG2}
\end{figure*}

There is a topological response function $\omega(T)$ based on the time-averaged expectation value of a geometric operator $\hat{O}^\lambda(t)$:
\begin{equation}
    \begin{aligned}
        \omega(T)=\frac{1}{\lambda^2 T}\int_0^T\langle \psi(t)|\hat{O}^\lambda (t)|\psi(t)\rangle dt
    \end{aligned}
\end{equation}
where $\hat{O}^\lambda (t):= 2g^{-1}(z_t)(p_{2, t}\partial_{x^1}\hat{H}-p_{1, t}\partial_{x^2}\hat{H})$, and $\lambda$ relates to the kinetic energy of the auxiliary particle (see Methods). Under the dual conditions of ergodicity ($T \to \infty$) and adiabaticity ($\lambda \to 0$), $\omega(T)$ converges to the first Chern number $C_1$.

We experimentally investigate this relationship using a superconducting qubit. Fig. \ref{fig:FiG1} presents the quantum state tomography and the extracted topological response for varying velocities $\lambda$. 
For a slow particle ($\lambda=1/20$), the driving is sufficiently adiabatic, and as the geodesic length increases, $\omega(T)$ converges to $C_1=1$, as shown in Fig. \ref{fig:FiG1}. 
Conversely, as the velocity increases, $\lambda=1/5$, the breakdown of adiabaticity introduces dynamic error. 
This result highlights the trade-off: exploring the manifold (ergodicity) requires long times, but doing so too quickly violates adiabaticity.

\emph{Ensemble-accelerated topological characterization.---}
To overcome the intrinsic limitations of single-particle trajectory tracking—most notably the pronounced sensitivity to initial conditions\cite{Strogatz2018nonlinear} that is characteristic of chaotic hyperbolic dynamics—we introduce a multi-particle ensemble approach. 
In classical chaotic systems, individual trajectories diverge exponentially from nearby initial conditions, leading to rapid loss of predictive power for single realizations and making long-time statistical predictions from single trajectories impractical. 
This sensitivity undermines the utility of single-trajectory analysis as a tool for extracting meaningful statistics\cite{Gaspard2005chaos, Casetti2000geometric}. 

To address this, we adopt an ensemble perspective rooted in ergodic theory, wherein the statistical properties of the dynamics can be inferred from a sufficiently large collection of independent trajectories rather than from a single path\cite{Nielsen2024}. 
According to ergodicity on hyperbolic manifolds, typical orbits explore the available phase space uniformly so that ensemble averages converge to the same value as time averages under appropriate conditions. 
This theoretical foundation justifies the use of ensemble equivalence and ensures that independent evolutions with distinct initial conditions yield statistically consistent measures of dynamical quantities in the long-time limit. 
By mapping each member of the ensemble to a separate quantum register and leveraging the intrinsic parallelism of quantum processors, our scheme permits simultaneous forward propagation of a large set of initial conditions. 
This allows us to construct a parallelization strategy
based on ensemble equivalence, enabling the efficient use
of computational resources to obtain statistically significant results within a shorter timeframe.

\begin{figure*}[t]
    \centering
    \includegraphics[width=17cm]{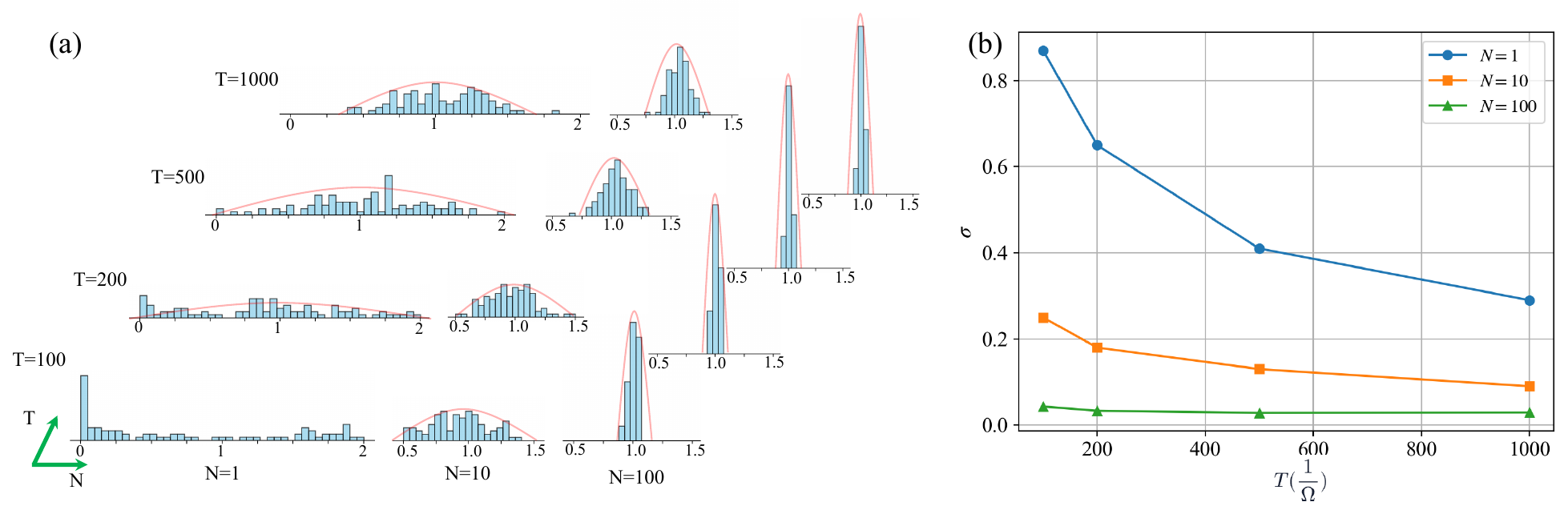}
    \caption{The relationship between the distribution and standard deviation of the dynamical response and the particle number and evolution time.
    With the increase in particle number and the extension of evolution time, the distribution of the dynamical response becomes more concentrated, and the standard deviation gradually decreases, converging to the first Chern number of the quantum system.}
    \label{fig:FiG3}
\end{figure*}

We distribute $N$ virtual particles uniformly on the Bolza surface according to the induced volume measure $\sqrt{|\det(g)|}dx^1\wedge dx^2$. 
They have the same initial speed but random directions.
The generalized topological response is averaged over the ensemble:
\begin{equation}
    \begin{aligned}
        \tilde{\omega}(T)=\frac{1}{N}\sum_{i=1}^{N} \frac{1}{\lambda_i^2 T}\int_0^T \langle \psi_i(t)|\hat{O}^{\lambda_i}_i (t)|\psi_i(t)\rangle.
    \end{aligned}
\end{equation}
Experimentally, we first use a numerical algorithm to generate a set of initial particle positions and propagation directions. 
The initial quantum state of each particle is determined by the local magnetic-field direction at its initial position. 
We then encode this information onto the quantum chip, allowing all qubits to evolve simultaneously.
Tomography results and convergence time for $N=10$ and $N=100$ are shown in Fig. \ref{fig:FiG2}. 
By engaging 10 and 100 qubits simultaneously to track independent geodesics, we observe a significant acceleration in convergence. 
As shown in Fig. \ref{fig:FiG2}(b), increasing $N$ reduces the time required for $\tilde{\omega}(T)$ to stabilize at the quantized value. This ensemble strategy effectively mitigates numerical instabilities inherent to long-time integration of chaotic systems while enhancing statistical significance.

Fig. \ref{fig:FiG3} shows the distribution of dynamical responses for different particle numbers and evolution times. As the particle number and time increase, the distribution of dynamical responses becomes more concentrated, and the standard deviation gradually decreases. This behavior indicates that the method we employ has good convergence and successfully achieves computational acceleration.

\begin{figure*}[t]
    \centering
    \includegraphics[width=17cm]{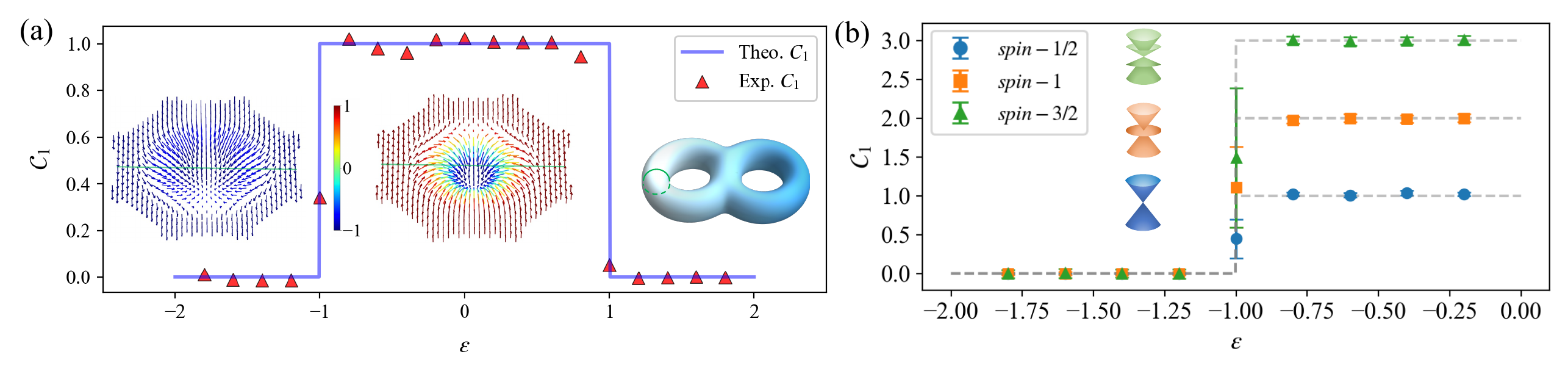}
    \caption{(a) The relationship between the first Chern number and $\epsilon$. The red triangles represent the converged values of $\tilde{\omega}(T)$, and the blue semi-transparent solid line represents the theoretical value of the first Chern number.
    Spin textures corresponding to the Hamiltonian polarization states at $\epsilon=0$ and $\epsilon=1.5$. The red lines indicate the gluing of the left and right boundaries, which forms a single handle of the double torus.
    (b) For different values of $\epsilon$, systems with different spins converge to their corresponding first Chern numbers.
    }
    \label{fig:FiG4}
\end{figure*}

\emph{Observation of topological phase transitions.---}
Here we demonstrate the universality of our method by probing topological phase transitions. By tuning the parameter $\epsilon$ in the Hamiltonian, we drive the system across distinct topological phases. Fig. \ref{fig:FiG4}(a) compares the experimentally extracted Chern number (red triangles) with theoretical predictions (blue line).

For $|\epsilon|<1$, the system resides in a topologically non-trivial phase, where the ground-state bundle corresponds to a generator of the cohomology group $H^2(\Sigma_2)$. Outside this range, the topology becomes trivial ($C_1=0$). The two spin-texture patterns plotted in Fig. \ref{fig:FiG4}(a) demonstrate that the system resides in two distinct topological phases.
The excellent agreement between experiment and theory confirms that the HDQS can accurately capture global topological invariants through local dynamical responses.

Extending the hyperbolically driven quantum system to accommodate higher-spin degrees of freedom—yielding the Chern number scaling $C_1 = 2n$—unlocks a rich landscape of underlying physics. Crucially, elevating the system from a two-level qubit to a spin-$1$ or higher-dimensional qudit transitions the internal symmetry from $SU(2)$ to $SU(N)$, enabling the synthesis of non-Abelian gauge fields and complex Berry connections within the negatively curved parameter space.  For example, particles with various spins\cite{Schwinger1965edited} can undergo polarization in hyperbolic space, and we can study different exotic material models—such as Maxwell metals\cite{Tan2018PRL} and Chern insulators\cite{Qi2011}—to uncover the unique properties they might exhibit in such spaces. These investigations allow hyperbolic manifolds to transcend being purely mathematical structures, connecting them instead to truly intriguing physical phenomena\cite{Berry1981quantizing}.
We extend our Hamiltonian by replacing the spin operators and numerically simulate the dynamics of spin-1 and spin-3/2 particles on a hyperbolic manifold in the presence of a magnetic field. 
The numerical results show good convergence. 
For $|\epsilon|<1$, the system is topologically nontrivial, yielding Chern numbers $C_1=2n$, where $n$ denotes the spin. Outside this parameter regime, the system becomes topologically trivial with $C_1=0$. 
We select some values of $\epsilon$ and plot the corresponding results in Fig. \ref{fig:FiG4}(b).

\emph{Discussion.---}
In this work, we have successfully simulated a hyperbolically driven quantum system (HDQS) on a superconducting circuit platform. By mapping the geometric properties of a Bolza surface onto the Hamiltonian parameters, we revealed how the intrinsic geometry of the parameter space governs the dynamical observables of the quantum system.

Our results elucidate the critical interplay between adiabaticity and ergodicity. While the Hopf–Anosov theorem ensures that single-particle geodesics are ergodic, the practical extraction of topological invariants is constrained by the need for adiabatic driving. We addressed this challenge by extending the framework to particle ensembles. This approach is conceptually analogous to the set-propagation method in chaos theory: since long-time numerical integration of chaotic geodesics is inherently unstable, short-time ensemble averaging provides a more robust and computationally efficient pathway.

Furthermore, this work demonstrates the unique advantage of quantum simulation in parameter space. Unconstrained by the Euclidean limitations of real space, our approach focuses on characterizing physical processes within parameter space to investigate their intrinsic properties. By leveraging the flexible controllability of superconducting qubits and the efficiency of parallel computing, this framework offers a novel avenue for exploring physical systems on exotic manifolds and provides fresh perspectives for manifold-based quantum machine learning.

\emph{Appendix.---}

\emph{Geodesic Dynamics.---}
The trajectory of the classical particle in the phase space $(z, p)=(x^1+ix^2, p_1+ip_2)\in \mathbb{C}^2$ is governed by the canonical equations of motion:
\begin{equation}
    \begin{aligned}
        \dot{x}^i = g^{ij}p_j, \quad \dot{p}_i = -\frac{1}{2}\frac{\partial g^{jk}}{\partial x^i}p_j p_k.
    \end{aligned}
\end{equation}
We numerically integrate these equations using a fourth-order Runge–Kutta method. The resulting trajectory $x(t)$ determines the time-dependent Hamiltonian $\hat{H}(x(t))$.

\emph{Experimental Implementation.---}
The experiment was conducted on a superconducting quantum processor with transmon qubits. The continuous time evolution operator $U(T)=\mathcal{P}\int_0^T e^{-i\hat{H}(x_t)t}dt$ was calculated by numerically integrating the Hamiltonian along the pre-computed geodesic. This unitary was then decomposed into a sequence of standard single-qubit gates and implemented on the hardware. Quantum state tomography was performed to reconstruct the density matrix at each time step.

The specific vector field used in the Hamiltonian is defined by:
\begin{equation}
    \begin{aligned}
        \begin{pmatrix}  \Omega_x \\ \Omega_y \\ \Omega_z \end{pmatrix}= 
        \Omega\begin{pmatrix} \cos f(z)(z + z^*)/2|z| \\ \cos f(z)(z - z^*)/2i|z| \\ \sin f(z)+\epsilon \end{pmatrix}
    \end{aligned}
\end{equation}
with the smoothing function $f(z)$ defined as:
\begin{equation}
    \begin{aligned}
        f(z) = \begin{cases} A \exp\left( -\frac{1}{\rho^{2} - |z|^{2}} \right) + B, & |z| < \rho \\ B, & |z| \ge \rho \end{cases}.
    \end{aligned}
\end{equation}
We chose parameters $A=\pi e^{1/\rho^2}$, $B=\pi/2$, and $\rho=0.6$ to ensure the smoothness of the mapping across the manifold.

\begin{figure*}[t]
    \centering
    \includegraphics[width=15cm]{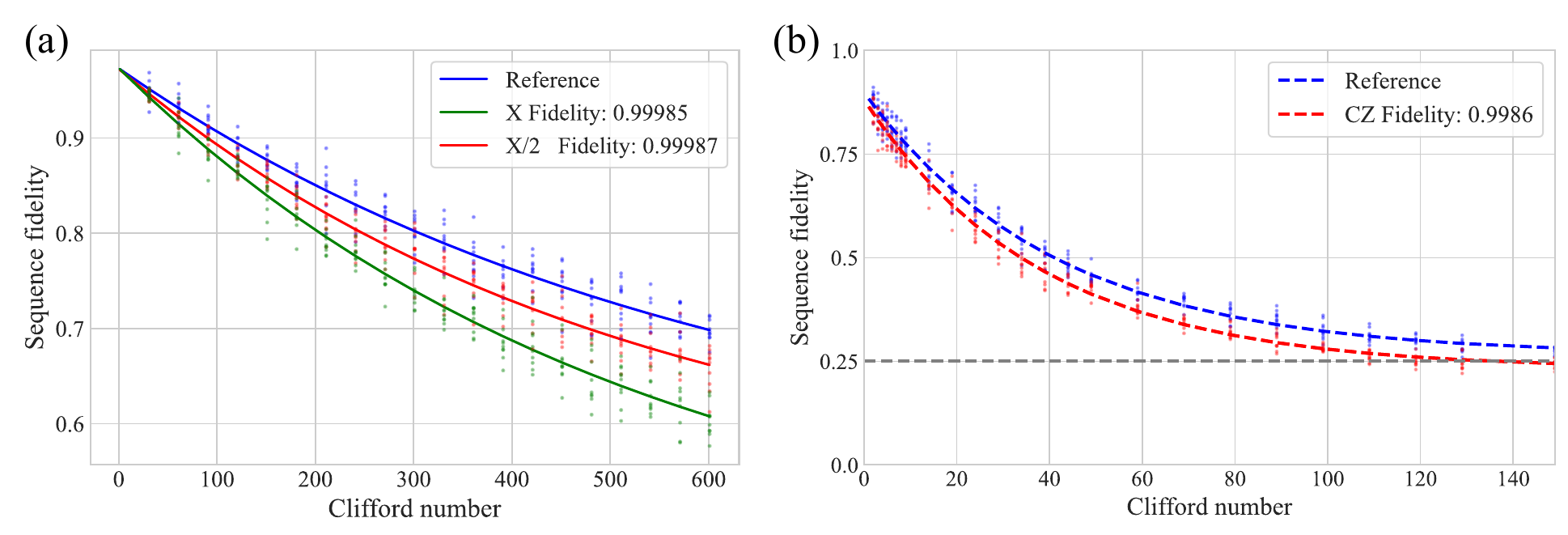}
    \caption{Interleaved randomized benchmarking}
    \label{fig:FiG5}
\end{figure*}

We characterize and optimize the gates with randomized benchmarking, of which results are shown in Fig.\ref{fig:FiG5}.

\begin{acknowledgements}
This work was supported by the Innovation Program for Quantum Science and Technology (Grant No.2021ZD0301702), NSFC (Grants No. U21A20436), NSF of Jiangsu Province (Grants No. BE2021015-1 and BK20232002), the Natural Science Foundation of Jiangsu Province (No. BK20233001), and the Natural Science Foundation of Shandong Province (Grant No. ZR2023LZH002), National Natural Science Foundation of China (Grants No. 92365206 and 92565301), 
The Innovation Program for Quantum Science and Technology (Grants No.2021ZD0301802).
\end{acknowledgements}

\bibliography{ref}

\end{document}